\documentclass[fleqn,usenatbib]{mnras}

\usepackage{newtxtext,newtxmath}
\usepackage[T1]{fontenc}

\DeclareRobustCommand{\VAN}[3]{#2}
\let\VANthebibliography\thebibliography
\def\thebibliography{\DeclareRobustCommand{\VAN}[3]{##3}\VANthebibliography}

\usepackage{graphicx}	
\usepackage{amsmath}	

\title[M-R relation of magnetized white dwarfs from SDSS]{Mass-Radius relation for magnetized white dwarfs from SDSS}

\author[Karinkuzhi, Mukhopadhyay, Wickramasinghe \& Tout]{
Drisya Karinkuzhi,$^{1}$\thanks{E-mail: drdrisyak@uoc.ac.in}
Banibrata Mukhopadhyay,$^{2}$\thanks{E-mail: bm@iisc.ac.in}
Dayal Wickramasinghe$,^{3}$
and Christopher A. Tout$^{4}$
\\
$^{1}$Department of Physics, University of Calicut, Malappuram. 673635, India\\
$^{2}$Department of Physics, Indian Institute of Science, Bangalore. 560012, India\\
$^{3}$Mathematical Sciences Institute, The Australian National University, Canberra, ACT 2601, Australia\\
$^{4}$Institute of Astronomy, The Observatories, Medingley Road, Cambridge CB3 OHA, UK
}

\date{Accepted 2024 March 18. Received 2024 February 29; in original form 2023 June 3}

\pubyear{2024}

\begin{document}
\maketitle

\begin{abstract}
We present the observational mass-radius (M–R) relation for a sample of 47 magnetized white dwarfs (WDs) with the magnetic field strength ($B$) ranging from 1 to 773 MG, identified from the SDSS data release 7 (DR7). We derive their effective temperature, surface gravity (log g), luminosity, radius,
and mass. While atmospheric parameters are derived using a Virtual Observatory Spectral Energy Distribution Analyzer
(VOSA), the mass is derived using their location in the HR diagram in comparison with the evolutionary
tracks of different masses. We implement this mass measurement instead of a more traditional method of deriving masses from
log g, which is unreliable as is based on SED and generates errors from other physical parameters involved. The main disadvantage
of this method is that we need to assume a core composition of WDs. As it is complicated to identify the exact composition
of these WDs from low-resolution spectra, we use tracks for the masses 0.2 to 0.4$M_\odot$ assuming a He-core, 
0.5 to 1.0 $M_\odot$ assuming CO core, and above $M_\odot$ assuming O-Ne-Mg core.
We compare the observed M–R
relation with those predicted by the finite temperature model by considering different $B$, which are well in agreement considering their relatively low surface fields,
$\lesssim 10^9$ G. Currently, there is no direct observational detection of magnetized WDs with $B > 10^9$ G.
We propose that our model can be further extrapolated
to higher $B$, which may indicate the existence of super-Chandrasekhar mass (M > $1.4M_\odot$) WDs at higher $B$.

\end{abstract}




\section{Introduction}
The number of highly magnetized white dwarfs (WDs), with a field strength between $10^3-10^9$ G, tremendously increased in the last few years, thanks to large field spectroscopic surveys. However, the role played by magnetic fields on stellar structure and evolution to the WD phase is still an open question. Moreover, the origin of this magnetic field is also actively debated. Magnetized WDs come in different varieties considering their photospheric nature, some are H-rich, called DA-type white dwarfs (DA-WDs) and some are He-enriched called DB-type white dwarfs (DB-WDs). Although most of the magnetized WDs are of DA type, there are a few WDs that show the enhancement of carbon in their atmosphere, the so-called DQ-WDs. A few of them also show weak metal lines in their atmospheres and are named DZ-WDs.
While the properties (masses and radii) of non-magnetic WDs can be determined with great accuracy using model atmospheres, these well-established techniques become increasingly less reliable, particularly for determining surface gravities as the surface field increases mainly because of the absence of reliable line-broadening theories (\citealt{Wickramasinghe2000,Ferrario2020,Kawka2018}). Our approach in this paper is the following. First, we use the effective temperatures from continuum spectra obtained from spectroscopic surveys of WDs with measured magnetic fields. Second, we consider the published theoretical cooling curves for different interior compositions to determine masses and radii initially assuming that the field only has a secondary effect on the cooling curves.


With these analyses, we also aim to understand and constrain the physical and chemical properties of the progenitors of over-luminous Type Ia supernovae (SNeIa). There are only a handful of over-luminous SNeIa detected to date (\citealt{Howell2006,Scalzo2010}) and their observational properties point towards the existence of super-Chandrasekhar ($M > 1.4 M_{\odot}$) WDs. Although the existence of super-Chandrasekhar WDs is not confirmed yet by the direct observations, it was already shown by our group  (\citealt{Upasana2012,Upasana2013,Subramanian2015}) that when we take into account $B$, modified Equation of State (EoS) of electron degenerate matter and/or Lorentz force arising due to magnetic field, along with rotation, lead to WDs with masses much higher than the Chandrasekhar-limit.  Recently, \citet{Abhay2020} considered a finite temperature model to predict the existence of super-Chandrasekhar WDs.  They obtained an interface between the electron degenerate-gas dominated inner core and the outer ideal gas surface layer or envelope by incorporating both the components of gas throughout the model WD.  With their model, they found that the Chandrasekhar Mass--Radius (M--R) relation can be retained in the absence of $B$, or the presence of a weak $B$, and if the temperature is kept constant in the core. However, at higher surface $B$ and considering central $B$ of the order of 10$^{14}$ G, a nonrotating WD mass could 
be  $\sim 2$ $M_\odot$. With a sample of moderately magnetized WD from Sloan Digital Sky Survey (SDSS) data release 7, we try whether we can satisfactorily reproduce their observed properties in accordance with this model so that it can be extrapolated to cases of the high surface $B$, in case they are observed in future.   

This paper is structured as follows. Section 2 describes the selection of the sample.  Subsequently, Sect. 3  discusses the method used for deriving the atmospheric parameters and physical quantities. In Sect. 4 we discuss the observational results in comparison with theoretical model predictions. Further, 
Sect. 5 attempts to identify the dependence of magnetic fields on different physical parameters. Conclusions are presented in Sect. 6. 

\section{Sample selection}
Our sample consists of magnetized WDs from the WD catalog of \citet{Kleinman2013} which is constructed based on the spectra from SDSS Data Release (DR) 7. \citet{Kepler2013} measured $B$ using the Zeeman splitting of Balmer lines for 521 WDs, which ranges from 1 to 773 MG. We select a sample of forty-seven (47) WDs from these 521 magnetized WDs, well separated by $B$. 

\begin{figure*}
\includegraphics[angle=0,height=16cm,width=20cm]{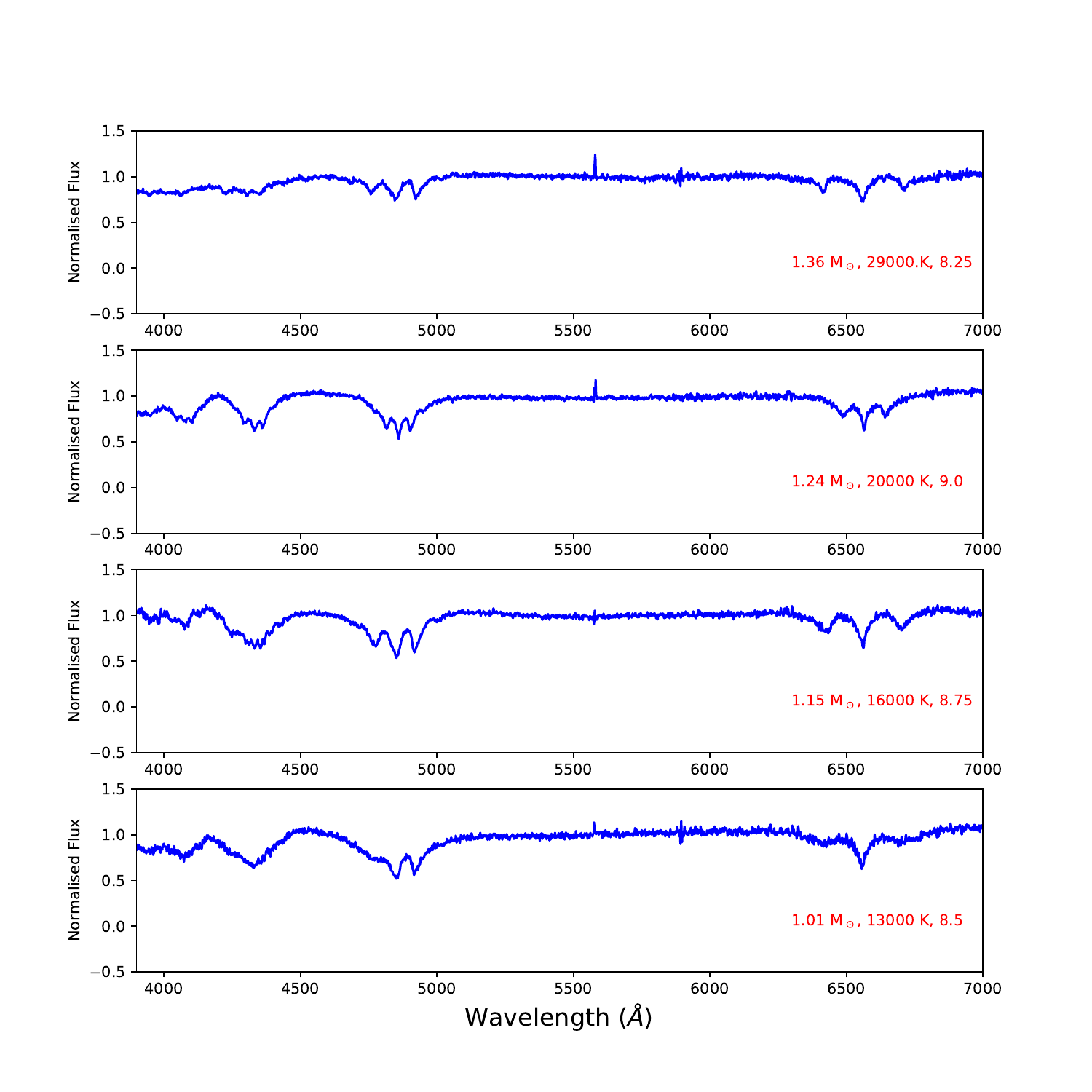}
      \caption{Sample spectra for the magnetized WDs from~\citet{Esteban2018} is presented. The mass, effective temperature, and log $g$, listed by \citet{Esteban2018},  are also presented in each panel.
\label{Fig:magnetic_GAIA}}
\end{figure*}

In Fig.~\ref{Fig:magnetic_GAIA}, we present a few sample spectra for massive ($>M_\odot$) WDs selected from \citet{Esteban2018} and flagged as magnetic WDs. These spectra are acquired from SDSS DR 16.  The presence of strong $B$ in these WDs is evident from the  Zeeman splitting of Balmer lines. What can be said with certainty based on the observed Zeeman splitting without detailed modeling is that the WD with the lowest field is that in the second panel from the top. The Zeeman splitting appears to be the largest in the upper panel, that could be due to the high $B$ in this specific WD. Also, we note here that this particular WD is relatively hotter compared to the other WDs in the figure. Others also have higher fields but how high is depends on the detailed modeling that must allow also for magnetic broadening caused by the field spread across the surface of the star (see \citealt{Wickramasinghe2000}).
Therefore, even though the parameters of these WDs vary significantly from each other, it is very difficult to find clear differences in their spectral characteristics visually. Hence, we apparently could not find any correlation among these spectra.  Moreover, the measurements of $B$ are also not available for them, so it is very difficult to draw a robust conclusion.

\section{Derivation of atmospheric parameters}
The atmospheric parameters (effective temperatures, T$_{eff}$, and surface gravity, log $g$) and luminosities for the sample WDs are measured with the help of the Spectral Energy Distribution (SED) tool, VOSA (Virtual Observatory SED Analyser; \citealt{Bayo2008}). VOSA is a Virtual Observatory tool that allows building the SEDs of objects using as many photometric points available from different photometric surveys ranging from ultraviolet to infrared, in an automated way. VOSA compares the photometric data with different collections of theoretical models and determines which model best reproduces the observed data following different statistical approaches. For this, magnitudes need to be converted to average fluxes first, adopting appropriate photometric zero points, and then compared with model fluxes averaged over the same filter bandpasses. We construct and analyze the SEDs for these WDs.  The atmospheric parameters are then estimated for each WD from the model \citep{Koester2010} that best fits the data. 

\begin{figure}
\includegraphics[angle=0,height=9cm,width=9cm]{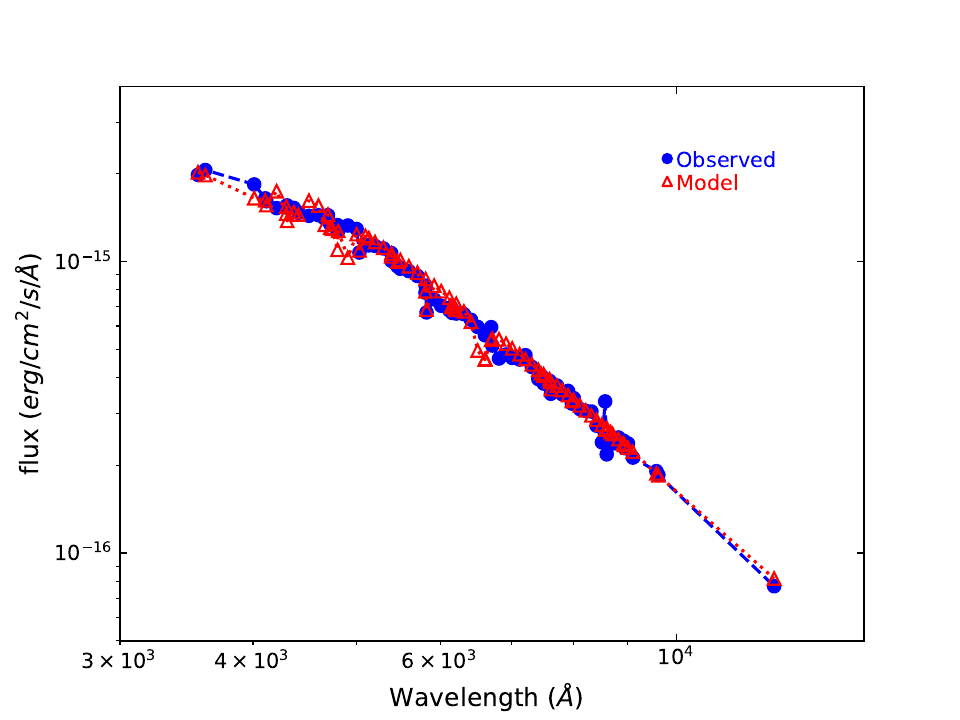}
\includegraphics[clip, trim=0cm 0cm 0cm 0cm,angle=0,height=9cm,width=9cm]{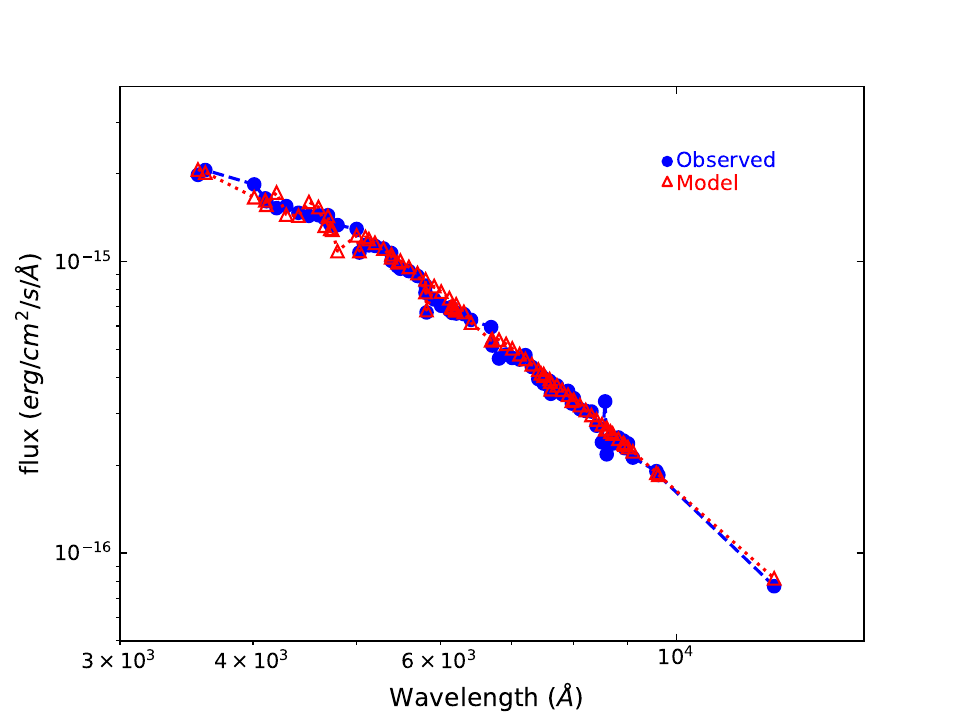}
      \caption{The SEDs are shown for the object RA: 207.92155, DEC: +54.32978. This object is the highest $B$ object in our sample. The upper panel shows the fit with the best model considering all the available data points till 12500 \AA. Although a few fluxes are available around 45000 \AA, we do not include them as these values are tagged as unreliable.  The temperature and log $g$ are found to be 11000 K and 9.25 respectively from this fit. The lower panel shows the best fit after removing the fluxes in the Balmer regions which is known to have been much affected by higher magnetic field. The derived temperature and log $g$  are 11250 and 9.5, respectively, in this case. An extinction value of 0.0317 is adopted in both cases. The temperature remains the same in both cases but log $g$ shows a slight deviation of 0.25. The fits shown here are reproduced using the data from the VOSA interface.   
\label{Fig:SED}}
\end{figure}

 However, we stress here the fact that the model we use is non-magnetic, hence our estimates might not be free from uncertainties since the presence of a strong magnetic field is expected to change the energy levels and subsequently the line strengths. As explained by \cite{Hardy2023a} for WD with hydrogen-rich envelopes, all the Balmer lines are severely affected. From Fig. 2 in \citealt{Hardy2023a}, it is clear that the H$_\alpha$ line shows a larger split in the presence of a stronger magnetic field than 100 MG. In our sample, however, only five WDs have $B$ $>$ 100 MG. To keep the uncertainties at a minimum, the parameters for these WDs are re-derived after removing the magnitudes from such bands that include the H$_\alpha$ line, to fit the SED.  We compare two SEDs for the object with the highest $B$ (773 MG) in our sample, presented in 
Fig. \ref{Fig:SED}.  The temperature derived from both the fits remains the same but a shift of 0.25 is noted for log $g$ values.  In this particular case, we have enough flux points at different regions to find a satisfactory fit, but for the WDs with less number of available fluxes at different wavelength bands, we expect an error of $\approx$ 250 K in temperature.   Hence we conclude that for the present sample with moderately higher fields, the uncertainties in parameters caused by the use of non-magnetic models are very small. Another possible source of error is the binary nature of the sample WDs. If the companion is close enough, there will be a non-negligible contribution for flux from the companion, thus resulting in an overestimation of atmospheric parameters. For the current sample of WDs, we do not have enough information on binarity and associated parameters. However, a detailed inspection of their spectra does not show an indication of a close companion.   

\begin{figure}
\includegraphics[angle=0,height=9cm,width=9cm]{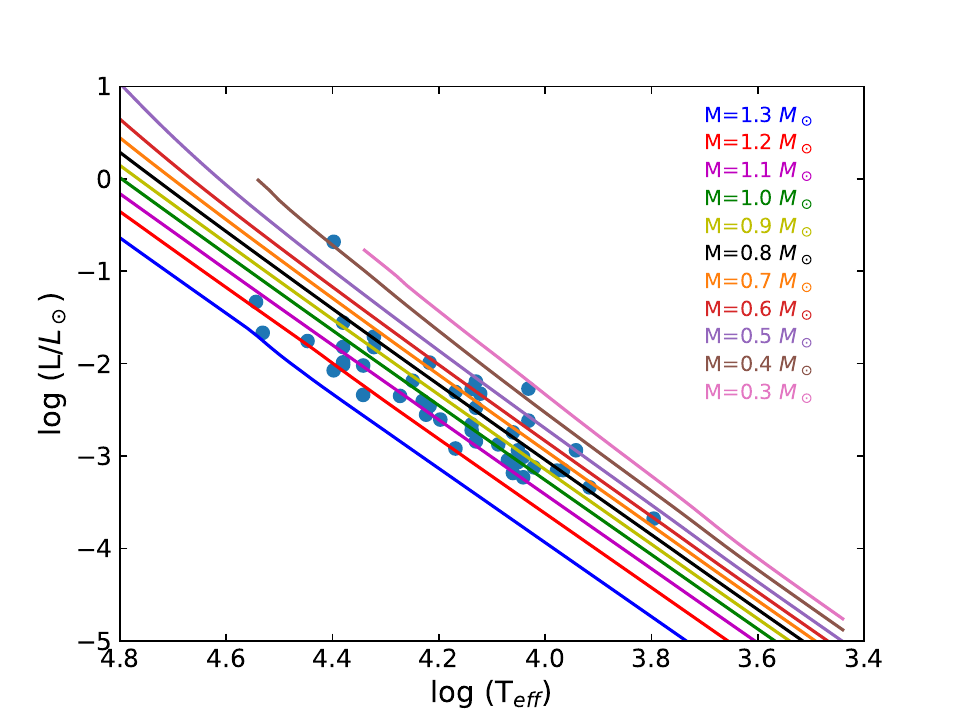}
      \caption{ Our sample WD data are in the HR diagram. The luminosity and T$_{eff}$ are derived using VOSA. The evolutionary tracks for different masses from \citet{Renedo2010} are also over-plotted for comparison.
\label{Fig:massevol}}
\end{figure}

The atmospheric parameters derived from the spectral fitting with appropriate models are known to be more reliable. However, we could not use them in our sample WDs because of their moderately high surface $B$ and associated Zeeman splitting. But, for those WDs with $B \le 30$ MG where the atmospheric structure is assumed to be not affected by field strengths, we have confirmed the results with a $\chi^2$ fitting method between the observed spectra and theoretical models. Thanks to accurate parallax from GAIA DR3, distances for these objects are derived. Since most of the WDs in our sample have known distances (D $\ge$ 100 pc), we consider the interstellar extinction ($A_v$) from \citet{Schlegel1998} to minimize the errors in luminosities. The maximum value of A$_v$ for the direction corresponding to the WD is assumed for the fit.  The most general method for deriving the mass of  WDs is based on their surface gravities. However, precise surface gravities are necessary for this measurement, otherwise, the uncertainty in log $g$ will directly reflect on the derived masses \citep{Schmidt1996}.  Here we would like to point out that for calculating the minimum $\chi^2$ value, VOSA uses the equation given below  
\begin{equation}
    {\chi_r}^2 = \frac{1}{N-n_p}\Sigma_{i=1}^{N}\left\{ \frac{(Y_{i,o} - M_d Y_{i,m})^2}{\sigma^2} \right\},
\end{equation}
where $N$ is the number of photometric points, $n_p$ is the number of fitted parameters for the model, $Y_{i,o}$ is the observed flux, $Y_{i,m}$ is the predicted flux,
$M_d$ is the multiplicative dilution factor, defined by $M_d$ = 
(${R}/{D})^2 $,
$R$ is the object's radius, and $D$ is the distance between the object and the observer.
The dilution factor is calculated based on best-fit too. Here it is clear that the accuracy of the result depends on the number of observed flux points and its error ($\sigma$). In order to obtain minimum ${\chi_r}^2$, the surface gravity (which is related to mass and radius) and effective temperature are varied until the best-fit values are obtained. The total observed flux $F_{obs}$ is calculated by integrating all the observed points after removing the overlap regions of the filter. Finally, bolometric luminosity is calculated using the equation L = $4\pi D^2 F_{obs}$. Although the radius and surface gravity are not explicitly used for calculating the luminosity, a smaller effect will be there through the dilution factor $M_d$. As the observed flux points are large for our sample, we assume that this effect is negligible. Moreover, surface gravity is a function of both the mass and the radius, and the uncertainty in its measurements will again affect the mass when it is measured using SED fitting. Hence, we estimate the mass of the WDs by finding the evolutionary track that passes through the corresponding luminosity and temperature in the  HR  diagram, given by Fig. \ref{Fig:massevol}. 
This method has been used by many authors (\citealt{Esteban2018,Provencal1998}) for deriving the mass for non-magnetized WDs. We use the evolutionary tracks of WDs for solar metallicity provided by \citet{Renedo2010}, those are constructed by assuming different core compositions. The available tracks for the masses 0.2 to 0.4 $M_\odot$ are calculated considering a He core. For higher masses, from 0.5 to 1.0 $M_\odot$, the tracks are constructed assuming a CO core and above $M_\odot$, an O-Ne-Mg core is assumed.  The derived masses and radii calculated from luminosities are also tabulated in Table~\ref{Tab:general}. 

\section{Interpretation of the results}
\begin{figure}
\includegraphics[angle=0,height=9cm,width=9cm]{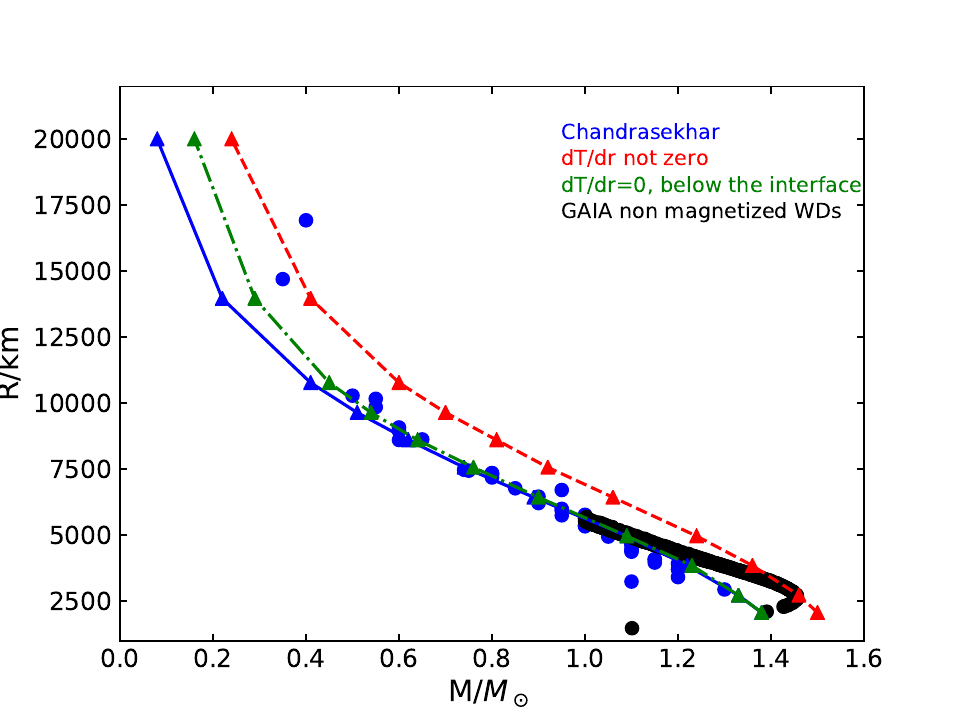}
      \caption{ WD M-R relation considering $dT/dr=0$ below the interface (green dot-dashed line), and $dT/dr \neq 0$ throughout (red dashed line), for L$=10^{-4}L_\odot$. Chandrasekhar's M-R relation (blue solid line) is also shown for comparison. Blue circles represent our sample WDs of $10^{-4}\lesssim{\rm L}/L_\odot\lesssim 10^{-1}$. Black circles are non-magnetized WD samples taken from \citet{Esteban2018}. 
\label{Fig:magnetic2}}
\end{figure}

We compare our results with the theoretical M--R relation predicted by \citet{Abhay2020}. The effect of the luminosities on the  M--R relation is being explored by these authors. The temperature inside WDs (more precisely in the core which extends close to the surface, up to the core-envelop interface) is generally assumed to be constant while deriving the M--R relation. However, \citet{Abhay2020} investigated the possibility of finite temperature gradient in the core, along with conventional uniform temperature, introducing the effect of luminosity with total pressure as the sum of the electron degenerate, ideal gas, and magnetic pressures. The electron degenerate pressure usually dominates over the ideal gas pressure in the stellar core and otherwise in the envelope.  An interface in the WD structure is defined to be the radius from where ideal gas pressure dominates the degenerate pressure till the surface.  
For the present purpose, we recollect some results of \citet{Abhay2020} for the sake of presentation and interpretation of data. 
In Fig. \ref{Fig:magnetic2} we compare our extracted M--R relation
from observed data
with those predicted by theoretical nonmagnetic models given by \citet{Abhay2020}. Non-magnetized WDs resulting from GAIA 100 pc WD sample \citep{Esteban2018} are also over-plotted for comparison.  It is clear from the figure that there is a shift from Chandrasekhar's original result (blue dashed line) for the M--R relation when considering a finite temperature gradient (red dashed line) in the core. Our objects extracted from data however seem to follow the curve corresponding to $dT/dr = 0$ in the core mostly, except for the small radius
and high mass ($\gtrsim M_\odot$) regime. We find that non-magnetized WDs from \citet{Esteban2018} follow the models corresponding to Chandrasekhar's M--R relation, except for higher masses and lower radii, similar to our objects. This may be due to the difference in compositions of WDs as these models are mainly constructed assuming CO core. More massive WDs are assumed to be having O-Ne-Mg core or iron core, though the existence of the latter is not confirmed yet. We further discuss this in Sect. \ref{Sect:composition}. Another important factor that decides the M--R relation is the thickness of the hydrogen envelope in the model. This effect has been extensively studied in literature \citep{Shipman1997,Provencal1998}
. \citet{Bergeron1992} found that the typical error in mass of WDs is at least 0.4 $M_\odot$ when considering a thin hydrogen envelope while it is originally thick. Moreover, for cooler WDs ($T_{eff}\le$ 12000 K), the presence of spectroscopically undetected helium in the atmosphere, which has been transported to the surface by convection, can also lead to an error in the derivation of masses while considering a hydrogen-rich model \citep{Bergeron1992}. 
\subsection{Effect of luminosity on the M--R relation}
\label{Sect:luminosity}
\begin{figure}
 \includegraphics[angle=0,height = 9cm,width=9cm]{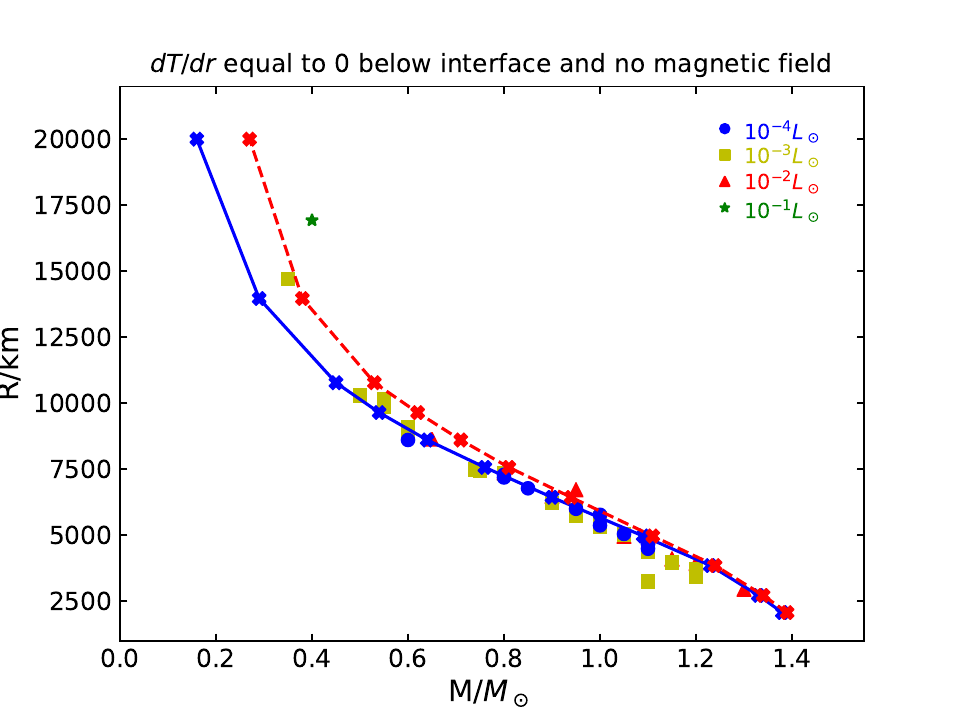}
    \caption{The M--R data is plotted for our WD candidates after grouping them based on their luminosities, where blue circles, yellow squares, red triangles, and green asterisks correspond to WDs with luminosities 10$^{-4}$, 10$^{-3}$, 10$^{-2}$ and 10$^{-1}$  $L_{\odot}$ respectively along 
    with theoretical curves denoted by solid blue and dashed red lines with crosses for respectively luminosities 10$^{-4}$  $L_{\odot}$ and 10$^{-2}$  $L_{\odot}$. 
\label{Fig:magnetic_lum}}
\end{figure}

To find the effect of luminosity on the M--R relation, we compare the results extracted from observed data with the theoretical M--R relation considering $dT/dr = 0$ below the interface and at different fixed luminosities. Blue solid and red dashed lines in Fig. \ref{Fig:magnetic_lum} represent the theoretical M--R relations derived for luminosities 10$^{-4}$ and 10$^{-2}$  $L_{\odot}$ respectively. Our objects are classified into four groups based on the measured luminosities, where blue circles correspond to WDs with luminosities 10$^{-4}$  $L_{\odot}$. The yellow squares, red triangles, and green asterisks represent WDs with luminosities 10$^{-3}$, 10$^{-2}$ and 10$^{-1}$  $L_{\odot}$ respectively. The observed M--R relations at various luminosities are well in agreement with theoretical predictions, except for
a few with L = $10^{-3}L_\odot$ in the high mass ($\gtrsim M_\odot$) regime.
\subsection{Effect of composition on the M--R relation}
\label{Sect:composition}
White dwarfs in the high mass tail of the WD mass distribution function are believed to have a core made up of Ne and Mg. As per \citet{Hamada61,Shapiro1983}, the zero-temperature configurations are known to be affected by internal composition. Moreover, the non-ideal contributions to the EoS of degenerate matter are dependent on the chemical composition, and these contributions are known to increase with the atomic number $Z$ of the chemical element. Similarly, for heavy elements like iron ($Z=26$ and $A = 56$), $\mu$ will be greater than 2. Hence for a fixed mass WD, the radius turns out to be a decreasing function of $Z$. Although the existence of WDs with heavy element core, like iron core, is still under debate, they are not completely ruled out (see \citealt{Isern1991}). Indeed,  the effect of chemical composition has been invoked by many authors \citep{Panei2000,Suh2000} in order to explain the deviation of a few WDs from the standard M--R relation. Here, in our Fig.~\ref{Fig:magnetic2}, at higher masses, a few WDs show a smaller radius than expected. We expect that this is due to the change in the composition of the core. These WDs also have large surface temperatures ($\approx$ 22000 K), which indicates a higher density of the core.
\subsection{Effect of magnetic field on the M--R relation}
\begin{figure}
 \includegraphics[angle=0,height = 9cm,width=9cm]{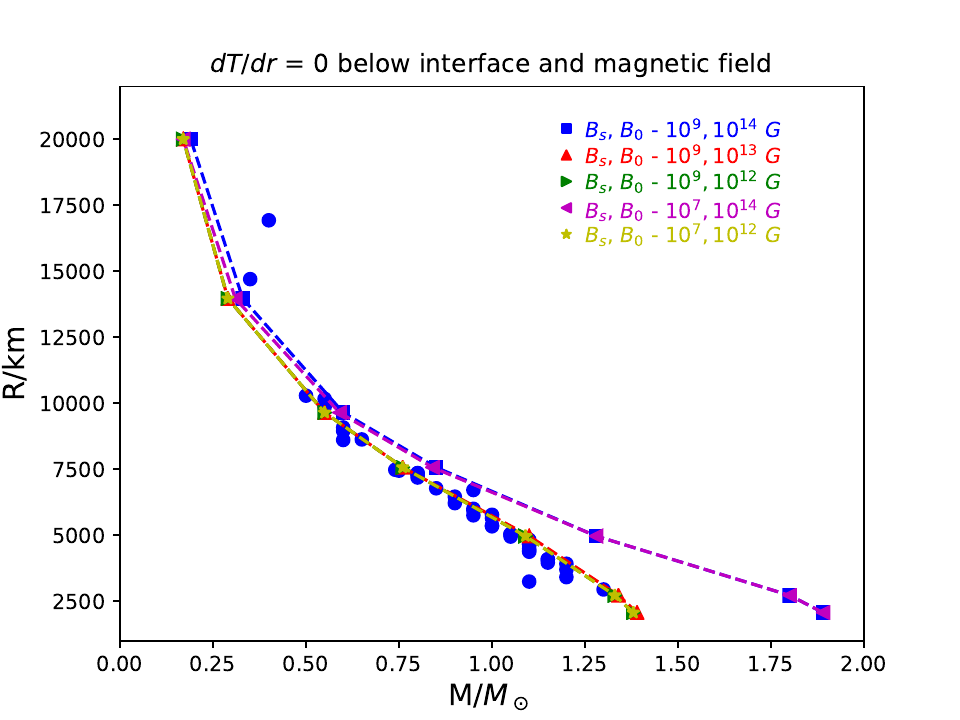}
    \caption{Same as in Fig. \ref{Fig:magnetic_lum} but the theoretical curves are constructed for different surface magnetic fields ($B_s$) and $B_0$, and at constant luminosity $10^{-4}$  $L_{\odot}$ with $dT/dr=0$ in the core. Different symbols represent different combinations of $B_s$ and $B_0$ as labeled in the figure. 
\label{Fig:magnetic_Mag}}
\end{figure}

In Fig. \ref{Fig:magnetic_Mag}, we include the effect of $B$ for the model corresponding to $dT/dr = 0$ in the core, but luminosity fixed at $10^{-4}$ $L_{\odot}$. The adopted $B$-s corresponding to different colors are labeled in the figure itself. The surface magnetic field ($B_s$) for our sample WDs is between 1 and 773 MG. There is a degeneracy between the M--R relations for $B_s$ of $10^7$ and $10^9$ G at lower central $B$ related to the parameter $B_0\sim 10^{12}$ and 10$^{13}$ G. For our modeling purpose, we 
assume density dependent field profile such that \\

\begin{figure}
\includegraphics[angle=0,height=9cm,width=9cm]{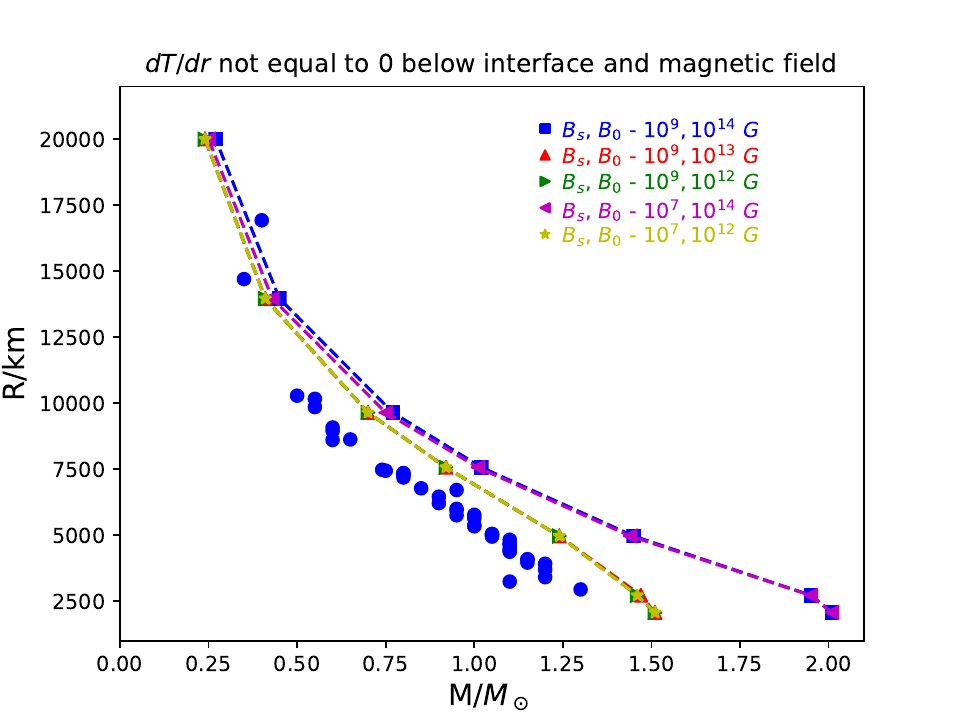}
      \caption{Same as in Fig.~\ref{Fig:magnetic_Mag}, but by considering $dT/dr \ne 0$.
\label{Fig:magnetic_Mag2}}
\end{figure}

$B \left(\frac{\rho}{\rho_{0}}\right)= B_{s} + B_{0}\left[1-exp\left (-\eta\left(\frac{\rho}{\rho_{0}}\right )^\gamma\right )\right]$,\\

\noindent where $\rho_{0}$ is a measure of reference density (chosen roughly 10 percent of the central matter density of the corresponding WD), $B_0$ is a parameter with the dimension of magnetic field and $\eta$ and $\gamma$ are dimensionless parameters which determine how exactly the field decays from the center to surface. 
Interestingly, our results extracted from data lie in the
M--R curve with lower $B_0$, as expected. Note that the M--R relations with lower $B_0$ practically do not change due to cooling and any further decay in fields, as shown by \cite{mukul2021}. We also present our results in  Fig. \ref{Fig:magnetic_Mag2} to compare them with the model 
$dT/dr \neq 0$ in the core, and at various $B$-s at the centre and the surface. There is a clear separation between the observed and model predictions when considering the theoretical M--R relation assuming $dT/dr \neq 0$. 

\section{Effect of magnetic field on the physical parameters of white dwarfs}
\begin{figure}
 \includegraphics[angle=0,height=9cm,width=9cm]{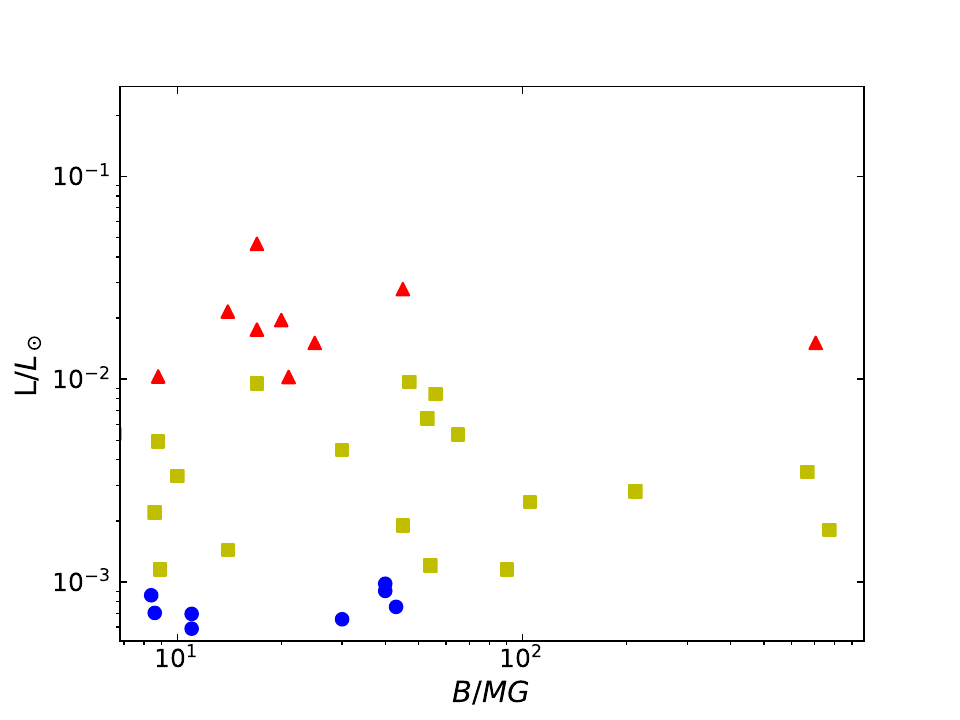}
   \caption{The derived Luminosity of WDs in our sample as a function of magnetic field strength ($B$). Symbols are the same as in Fig. \ref{Fig:magnetic_lum}.
\label{Fig:magnetic_LB}}
\end{figure}

\begin{figure}
 \includegraphics[angle=0,height =9cm,width=9cm]{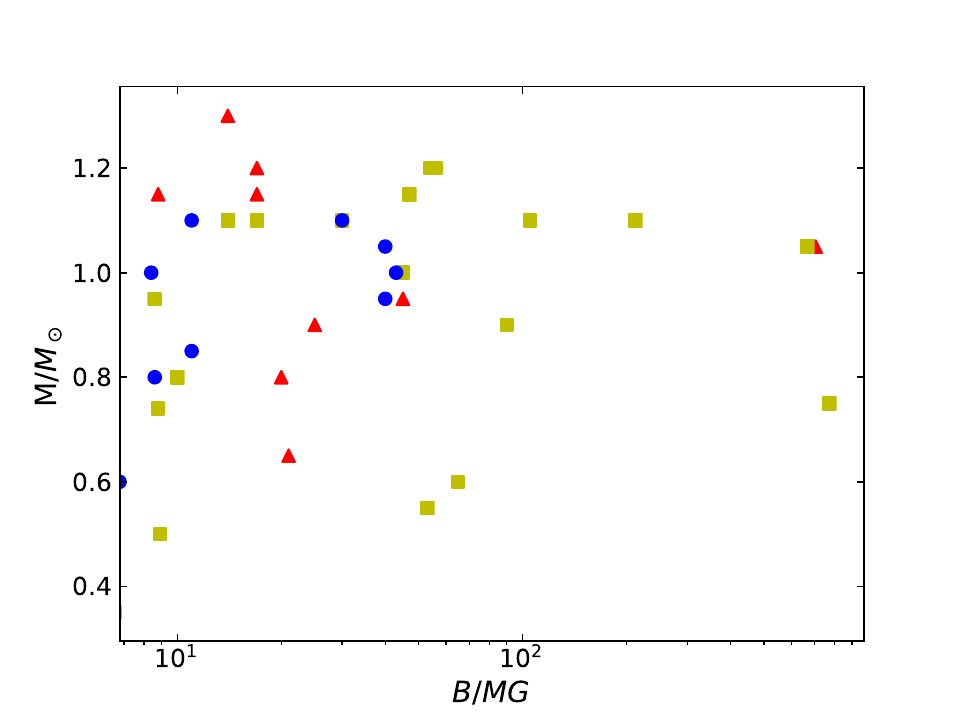}
   \caption{The mass of sample WDs as a function of $B$. 
\label{Fig:magnetic_MB}}
\end{figure}

\begin{figure}
 \includegraphics[angle=0,height = 9cm,width=9cm]{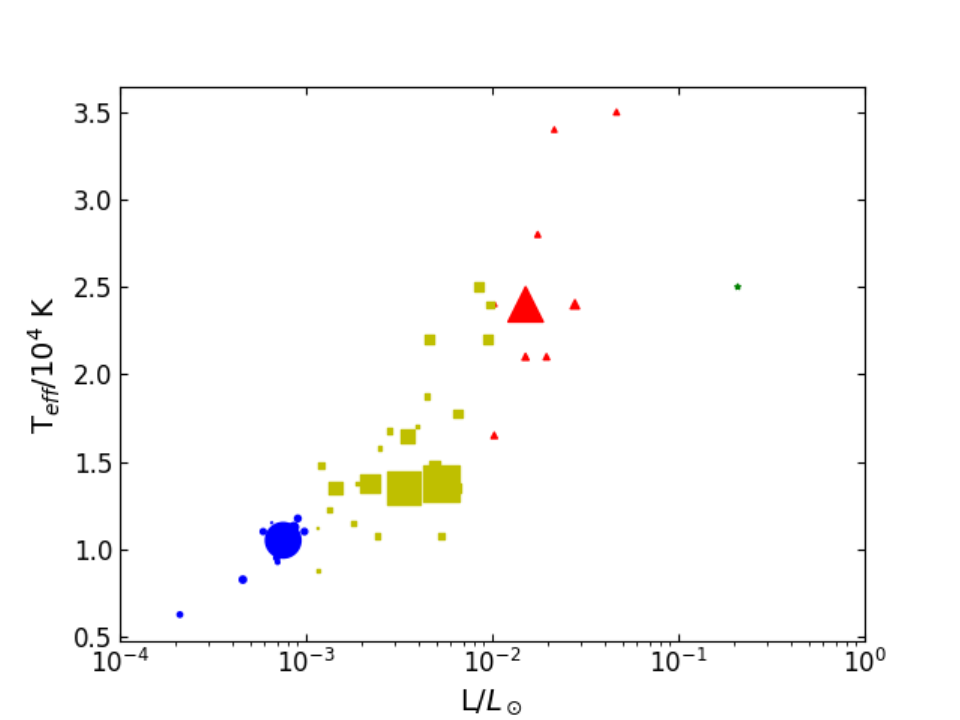}
    \caption{Luminosity of the same sample WDs given by
    Fig. \ref{Fig:magnetic_LB}  as a function of their effective temperature (T$_{eff}$) and the points are scaled to their $B$.
\label{Fig:magnetic_TL}}
\end{figure}

This section examines the effect of $B$ on the Luminosity and the mass of WD samples. Understanding the physical ($T_{eff}$, log $g$, $M$, $R$) and chemical properties of the WDs is important to understand their evolution. The properties of non-magnetized WDs have been studied extensively (see \citealt{Bedard2017} and references therein). For magnetized WDs, the situation is more complex as strong central $B$ is likely to modify these properties. Figures \ref{Fig:magnetic_LB} and \ref{Fig:magnetic_MB} present the luminosity and mass as functions of $B$.  Our previous theoretical analyses established the effect of $B$ on luminosities. Since there is no nuclear energy generation inside the WDs, the luminosity is expected to be constant and is equal to the corresponding surface luminosity. In the case of strong $B$, it is not wrong to assume that convection will be inhibited which in turn supports the assumption of constant luminosities. From Fig. \ref{Fig:magnetic_LB}, we could not see a clear trend of decreasing luminosity with stronger $B$, unlike what was predicted by \citet{Abhay2020}. Nevertheless, it is indicated from Fig. \ref{Fig:magnetic_TL} that the increase in luminosity is due to the increase in surface temperature. To be more clear, in Fig. \ref{Fig:magnetic_TL}, we scale the points according to their $B$. We note here that some WDs with lower $B$ have comparatively higher temperatures and higher luminosities. But for the highest $B$ WDs in our sample, although temperatures are lower compared to most of the sample WDs, their luminosities are at a moderate level.  Hence, Fig. \ref{Fig:magnetic_LB} combined with Fig. \ref{Fig:magnetic_TL} clearly indicates that there is no correlation of field with the cooling age of the magnetic WDs. 
Indeed, highly magnetized WDs are expected to be massive compared to their non-magnetic counterparts \citep{Ferrario2015,Ferrario2020}. However, we note from Fig. \ref{Fig:magnetic_MB} that the highest mass WDs are not the ones with the highest $B$. This could be simply due to their lower central $B$ -- their strength is not enough to modify the stellar structure. Also, as shown by \citet{Subramanian2015}, highly magnetized WDs with poloidally dominated fields are generally found to be lighter compared to toroidally dominated ones. 

\begin{table*}
\caption{Basic parameters for the magnetic WDs: T$_{eff}$, log $g$ and  L/$L_\odot$ are derived using VOSA, and mass M is derived from their location in the HR diagram using the cooling tracks. By using non-magnetic models, we expect to incur an error of approximately 250 K in temperature and 0.25 in log $g$, though the best fit shown in Fig. \ref{Fig:SED}, for the highest $B$ WD in our sample, does not show any shift in temperature. The typical uncertainties in temperature, caused by other uncertainties such as the number of flux points and uncertainty in their measurements,  when derived from the SED fitting, are presented along with the temperature values. Similarly, the uncertainties in log $g$ are $\pm$ 0.13 for all the values.  The error in luminosities is given in column $\delta$ L/$L_\odot$. This error leads to uncertainties in M and R. Only a small number of WDs have large errors, but they are not the ones with high $B$. The use of non-magnetic models is therefore not associated with significant uncertainties. Although the errors in the temperatures and luminosities are presented, we do not estimate the uncertainties in mass as it is measured from the locations of WDs in the H-R diagram.  }  
\label{Tab:general}
\begin{tabular}{cclccllllll}
\hline
\\
RA  &  DEC & Distance $\pm$ $\delta$d (pc)&  T$_{eff}$$\pm$ $\delta$ T$_{eff}$(K) &log $g$ $\pm$ 0.13 & L/$L_\odot$& $\delta$ L/$L_\odot$ & R ($R_\odot$)&$\delta R (R_\odot)$ & M ($M_\odot$)&$B$ \\
Degrees &  Degrees& (pc)  &  (K) &CGS & &  & & &&(MG) \\
\hline\\
002.64570   & 	+24.85873   & 	193.12 $\pm$ 14.24  & 	11000 $\pm$ 125    & 	8.00     &    5.883e-4   & 	2.398e-4   & 	6.675e-3   & 1.369e-3  & 1.1    &    11 \\
025.62735   & 	+00.58410   & 	635.86 $\pm$ 204.51  & 	10750 $\pm$ 125  & 	7.25   &    5.364e-3   & 	5.593e-3   & 	2.111e-2   & 1.101e-2  &0.35   &  6.6\\
032.95074   & 	+21.26345   & 	59.46 $\pm$ 0.96   & 	15750 $\pm$ 125  & 	9.00      &    2.480e-3   & 	1.896e-4   & 	6.685e-3   & 2.767e-4   &1.1    &  105\\
039.03908   & 	-08.13995   & 	190.54 $\pm$ 1.58   & 	14750 $\pm$ 125  & 	8.25   &    1.205e-3   & 	3.345e-4   & 	5.313e-3   & 7.431e-4   &1.2    &  54   \\
051.61739   & 	+05.36009   & 	308.32 $\pm$ 24.59  & 	35000 $\pm$ 125  & 	6.50    &    4.651e-2   & 	7.866e-3   & 	5.863e-3   & 5.233e-4   &1.15   &  17\\
102.12008   & 	+84.06141   & 	252.09 $\pm$ 17.79  &  12250 $\pm$ 125  & 	8.75   &    1.335e-3   & 	3.793e-4   & 	8.108e-3   & 1.164e-3   &1.0    &  4.3\\
118.14573   & 	+17.42364   & 	112.39 $\pm$ 2.06  & 	9500 $\pm$ 125   &   8.00      &    6.957e-4   & 	9.552e-5   & 	9.732e-3   & 7.156e-4   &0.85   &  11\\
119.56934   & 	+12.24132   & 	268.28 $\pm$ 22.70   & 	17750 $\pm$ 125  & 	6.50    &    6.519e-3   & 	1.209e-3   & 	8.534e-3   & 8.006e-4   &0.95   &  4.4\\
121.93056   & 	+39.64144   & 	514.70 $\pm$ 185.30  &  13750 $\pm$ 125   & 	8.25   &    5.320e-3   & 	4.226e-3   & 	1.285e-2   & 5.108e-3   &0.6    &  65\\
122.90137   & 	+46.19902   & 	1545.82 $\pm$ 697.82 &  25000 $\pm$ 125   & 	8.00      &    2.082e-1   & 	3.478e-1   & 	2.431e-2   & 2.033e-2   &0.4    &  3.6\\
130.50590   & 	+15.66164   & 	404.26 $\pm$ 76.94  & 	28000 $\pm$ 125  & 	6.50    &    1.755e-2   & 	7.191e-3   & 	5.627e-3   & 1.170e-3   &1.2    &  17\\
133.28787   & 	+56.57821   & 	163.73 $\pm$ 9.60  & 	8250 $\pm$ 125   &   7.50    &    4.577e-4   & 	1.176e-4   & 	1.047e-2   & 1.382e-3   &0.8    &  1\\
134.20703   & 	+25.57810   & 	92.17 $\pm$ 1.08   & 	11250 $\pm$ 125  & 	8.75   &    1.151e-3   & 	1.182e-4   & 	8.925e-3   & 4.996e-4   &0.9    &  90\\
142.85894   & 	+32.32947   & 	240.81 $\pm$ 17.647  & 	13750 $\pm$ 125  & 	7.25   &    2.199e-3   & 	5.912e-4   & 	8.261e-3   & 1.120e-3   &0.95   &  8.6\\
143.56656   & 	+29.75013   & 	345.21 $\pm$ 37.78  & 	21000 $\pm$ 125   & 	6.50    &    1.509e-2   & 	3.751e-3   & 	9.277e-3   & 1.235e-3   &0.9    &  25\\
145.64594   & 	+20.86895   & 	195.68 $\pm$ 8.06  & 	22000 $\pm$ 125  & 	6.50    &    9.533e-3   & 	9.064e-4   & 	6.718e-3   & 4.419e-4   &1.1    &  17\\
146.79631   & 	+00.63185   & 	281.95 $\pm$ 38.65  &  17000 $\pm$ 125  & 	6.50    &    3.943e-3   & 	1.287e-3   & 	7.235e-3   & 1.185e-3   &1.05   &  3.6\\
150.98459   & 	+05.64046   & 	154.03 $\pm$ 3.93  & 	16500 $\pm$ 125   & 	6.50    &    3.479e-3   & 	3.313e-4   & 	7.215e-3   & 3.605e-4   &1.05   &  668\\
154.52101   & 	+01.18987   & 	49.39 $\pm$ 0.17  & 	11000 $\pm$ 125  & 	9.00      &    9.799e-4   & 	2.191e-4   & 	8.615e-3   & 9.830e-4   &0.95   &  40\\
164.11866   & 	+65.38707   & 	510.61 $\pm$ 80.46  & 	16500 $\pm$ 125  & 	8.00      &    1.025e-2   & 	4.030e-3   & 	1.239e-2   & 2.441e-3   &0.65   &  21\\
170.12642   & 	-11.84754   & 	202.11 $\pm$ 12.17  & 	24000 $\pm$ 500  & 	8.50    &    1.031e-2   & 	1.747e-3   & 	5.871e-3   & 5.542e-4   &1.15   &  8.8\\
175.02662   & 	+61.16893   & 	472.87 $\pm$ 65.38  & 	13500 $\pm$ 125   & 	7.25   &    6.388e-3   & 	2.513e-3   & 	1.460e-2   & 1.622e-1   &0.55   &  53\\
181.54098   & 	+08.22326   & 	177.27 $\pm$ 8.19  & 	11750 $\pm$ 125   & 	8.00      &    9.030e-4   & 	4.034e-4   & 	7.248e-3   & 2.885e-3   &1.05   &  40\\
184.14736   & 	-00.44898   & 	252.01 $\pm$ 25.74  & 	13750 $\pm$ 125  & 	6.50    &    1.894e-3   & 	7.420e-4   & 	7.665e-3   & 1.508e-3   &1.00   &  45\\
185.70473   & 	+48.19252   & 	136.62 $\pm$ 3.00  & 	9250 $\pm$ 125   &  7.75    &    7.034e-4   & 	9.215e-5   & 	1.032e-2   & 7.314e-4   &0.80   &  8.6\\
188.01748   & 	+52.43007   & 	186.19 $\pm$ 6.05   & 	8750 $\pm$ 125    & 7.50      &    1.153e-3   & 	1.649e-4   & 	1.477e-2   & 1.137e-3   &0.50   &  8.9\\
192.21380   & 	-02.49021   & 	218.29 $\pm$ 8.55  & 	14750 $\pm$ 125   & 	7.00      &    4.925e-3   & 	6.650e-4   & 	1.074e-2   & 7.477e-4   &0.75   &  8.8 \\
193.97245   & 	+15.43197   & 	527.82 $\pm$ 86.13  & 	21000 $\pm$ 125  & 	6.50    &    1.956e-2   & 	8.448e-3   & 	1.056e-2   & 2.335e-3   &0.8    &  20\\
195.13940   & 	+59.06856   & 	64.31 $\pm$  1.92  & 	6250 $\pm$ 125   &   9.50    &    2.101e-4   & 	4.910e-5   & 	1.236e-2   & 1.526e-3   &0.6    &  6.8\\
203.41811   & 	+64.10760   & 	100.75 $\pm$ 1.03  & 	13500 $\pm$ 125   & 	8.50    &    1.440e-3   & 	1.508e-4   & 	6.934e-3   &3.850e-4    &1.1    &  14\\
207.92155   & 	+54.32978   & 	70.79 $\pm$ 0.12   & 	11500 $\pm$ 125   & 	8.75   &    1.803e-3   & 	3.857e-4   & 	1.069e-2   & 1.167e-3   &0.75   &  773\\
207.08668   & 	+38.17143   & 	139.24 $\pm$ 2.24   & 	34000 $\pm$ 500   & 	6.50    &    2.154e-2   & 	6.556e-4   & 	4.228e-3   &1.973e-4    &1.3    &  14\\
216.76404   & 	+37.35294   & 	111.45 $\pm$ 1.12  & 	18750 $\pm$ 125  & 	9.00      &    4.470e-3   & 	1.251e-4   & 	6.333e-3   & 1.224e-4   &1.1    &  30\\
227.87566   & 	+42.33970   & 	97.27 $\pm$ 0.86   & 	11250 $\pm$ 125   & 	8.25   &    8.595e-4   & 	1.035e-4   & 	7.714e-3   & 4.952e-4   &1.0    &  8.4\\
233.45428   & 	+00.98784   & 	204.19 $\pm$ 6.18   & 	13250 $\pm$ 125   & 	6.50    &    4.728e-3   & 	7.661e-4   & 	1.304e-2   & 1.085e-3   &0.6    &  2\\
234.62205   & 	+53.10129   & 	294.87 $\pm$ 18.75  & 	13500 $\pm$ 125  & 	8.25   &    3.335e-3   & 	8.052e-4   & 	1.055e-2   &1.289e-3    &0.8    &  10\\
240.99143   & 	+14.15827   & 	105.23 $\pm$ 1.59  & 	10500 $\pm$ 125   & 	9.50    &    7.535e-4   & 	6.682e-5   & 	8.291e-3   &4.173e-4    &1.0    &  43\\
256.00002   & 	+32.22461   & 	384.31 $\pm$ 81.25  & 	25000 $\pm$ 500  & 	6.75   &    8.457e-3   & 	4.950e-3   & 	4.900e-3   &1.447e-3    &1.2    &  56\\
257.31817   & 	+23.68650   & 	311.53 $\pm$ 47.02  & 	22000 $\pm$ 500   & 	7.50    &    4.582e-3   & 	1.622e-3   & 	4.658e-3   &8.511e-4    &1.1    &  6\\
260.87144   & 	+54.13216   & 	116.33 $\pm$2.15  & 	11500 $\pm$ 125   & 	8.50    &    6.545e-4   & 	1.276e-4   & 	6.442e-3   &6.434e-4    &1.1    &  30\\
263.14665   & 	+59.09262   & 	240.20 $\pm$ 7.84  & 	10750 $\pm$ 125   & 	7.75   &    2.412e-3   & 	2.651e-4   & 	1.415e-2   &8.446e-4    &0.55   &  1\\
327.37809   & 	-07.46997   & 	189.82 $\pm$ 5.12  & 	24000 $\pm$ 500   & 	7.00      &    2.779e-2   & 	2.398e-3   & 	9.638e-3   & 5.781e-4   &0.95   &  45\\
334.61910   & 	-00.00338   & 	121.74 $\pm$ 2.51  & 	16750 $\pm$ 125  & 	6.50    &    2.801e-3   & 	2.224e-4   & 	6.281e-3   & 2.665e-4   &1.1    &  212\\
341.92274   & 	+14.94412   & 	102.36 $\pm$ 1.22  & 	24000 $\pm$ 500  & 	7.50    &    9.691e-3   & 	3.796e-4   & 	5.691e-3   &2.620e-4    &1.15   &  47\\
356.52269   & 	+38.89381   & 	240.03 $\pm$ 12.90  & 	24000 $\pm$ 125   & 	6.50    &    1.511e-2   & 	2.075e-3   & 	7.106e-3   &5.707e-4    &1.05   &  706 \\

 \hline\\
\end{tabular}

\end{table*}

\section{Conclusions}

The mass-radius relation for a sample of magnetized WDs selected from SDSS DR7 is extracted. Instead of the traditional methodology for deriving the mass from the surface gravities, we have used the location of WDs in the HR diagram and then compared it with cooling tracks to derive their masses. The main source of error in this methodology is the assumption of the core mass. As it is difficult to understand the composition of the WDs from the low-resolution spectra, we have taken the tracks corresponding to O-Ne-Mg core for mass greater than $M_{\odot}$, and CO core for the masses 0.5 to 1.0 $M_{\odot}$.  We then have compared our results with theoretical predictions from the finite temperature model including the effect of different surface magnetic field strengths. We could get a satisfactory fit of observation with theory at lower surface fields. However, our model predicts super-Chandrasekhar WDs at higher fields. Nevertheless, there is no observational detection of super-Chandrasekhar WDs yet.

\section*{Acknowledgements}

The authors thank Abhay Gupta for providing some numerical results reported in an earlier theoretical paper to plot along with the results extracted from observation. Thanks are also due to the anonymous referee for carefully reviewing the manuscript with constructive comments, which have made the paper better. DK acknowledges the financial support from CSIR-India
through file No.13(9086-A)2019-Pool. BM acknowledges partial support
by a project of the Department of Science and Technology 
(DST-SERB) with research Grant No. DSTO/PPH/BMP/1946
(EMR/2017/001226). CAT thanks Churchill College for his fellowship. This publication makes use of VOSA, developed under the Spanish Virtual Observatory (https://svo.cab.inta-csic.es) project funded by MCIN/AEI/10.13039/501100011033/ through grant PID2020-112949GB-I00.
VOSA has been partially updated by using funding from the European Union's Horizon 2020 Research and Innovation Programme, under Grant Agreement nº 776403 (EXOPLANETS-A).

\section*{Data Availability}

Any new data generated during the work will
be made available whenever required by the readers.



\bibliographystyle{mnras}
\bibliography{WD-ref.bib} 




\appendix

\section{Some extra material}

If you want to present additional material that would interrupt the flow of the main paper,
it can be placed in an Appendix which appears after the list of references.


\bsp	
\label{lastpage}
\end{document}